\documentclass[twocolumn]{bmcart}

\usepackage{amsthm,amsmath}
\RequirePackage{hyperref}
\usepackage[utf8]{inputenc} 

\usepackage{graphicx}

\startlocaldefs
\endlocaldefs

\begin{document}

\begin{frontmatter}

\begin{fmbox}
\dochead{Method}

\title{The design and construction of reference pangenome graphs}

\author[
   addressref={aff1,aff2},          
   corref={aff1,aff2},              
   email={hli@ds.dfci.harvard.edu}  
]{\inits{HL}\fnm{Heng} \snm{Li}}
\author[
   addressref={aff1,aff2},
]{\inits{XF}\fnm{Xiaowen} \snm{Feng}}
\author[
   addressref={aff2},
]{\inits{CC}\fnm{Chong} \snm{Chu}}

\address[id=aff1]{
  \orgname{Department of Data Sciences, Dana-Farber Cancer Institute}, 
  \city{Boston, MA 02215},                    
  \cny{USA}                                   
}
\address[id=aff2]{%
  \orgname{Department of Biomedical Informatics, Harvard Medical School},
  \city{Boston, MA 02215},
  \cny{USA}
}

\begin{abstractbox}

\begin{abstract} 
The recent advances in sequencing technologies enables the assembly of
individual genomes to the reference quality. How to integrate multiple genomes
from the same species and to make the integrated representation accessible to
biologists remain an open challenge. Here we propose a graph-based data model
and associated formats to represent multiple genomes while preserving the
coordinate of the linear reference genome. We implemented our ideas in the
minigraph toolkit and demonstrate that we can efficiently construct a pangenome
graph and compactly encode tens of thousands of structural variants missing
from the current reference genome.
\end{abstract}

\begin{keyword}
\kwd{bioinformatics}
\kwd{genomics}
\kwd{pangenome}
\end{keyword}

\end{abstractbox}

\end{fmbox}

\end{frontmatter}

\section*{Background}

The human reference genome is a fundamental resource for human genetics and
biomedical research. The primary sequences of the reference genome
GRCh38~\cite{Schneider:2017aa} are a mosaic of haplotypes with each haplotype segment derived
from a single human individual. They cannot represent the genetic diversity in
human populations and as a result, each individual may carry thousands of large
germline variants absent from the reference genome~\cite{Huddleston:2017aa}.
Some of these variants are likely associated with phenotype~\cite{Eichler_2010}
but are often missed or misinterpreted when we map sequence data to GRCh38, in
particular with short reads~\cite{Li:2018aa}. This under-representation of
genetic diversity may become a limiting factor in our understanding of genetic
variations.

Meanwhile, the advances in long-read sequencing technologies make it possible
to assemble a human individual to a quality comparable to
GRCh38~\cite{Schneider:2017aa,Wenger_2019}. There are already a dozen of
high-quality human assemblies available in GenBank~\cite{Audano:2019aa}.
Properly integrating these genomes into a reference \emph{pangenome}, which
refers to a collection of genomes~\cite{cpgc:2016aa}, would potentially address
the issues with a single linear reference.

A straightforward way to represent a pangenome is to store unaligned genomes
in a full-text index that compresses redundancies in sequences identical
between individuals~\cite{Makinen:2010aa,Liu_2016,Boucher_2019}. We may
retrieve individual genomes from the index, inspect the k-mer spectrum and test
the presence of k-mers using standard techniques. In principle, it is also
possible to apply canonical read alignment algorithms to map sequences to
the collection, but in practice, the redundant hits to multiple genomes will
confuse downstream mapping-based analyses~\cite{NA2016159}. It is not clear how
to resolve these multiple mappings.

The other class of methods encodes multiple genomes into a sequence graph,
usually by collapsing identical or similar sequences between genomes onto a
single representative sequence. The results in a \emph{pangenome graph}. A
pangenome graph is a powerful tool to identify core genome, the part of a
genome or gene set that is shared across the majority of the strains or related species
in a clade~\cite{Vernikos:2015aa}. A common way to construct a basic pangenome
graph is to generate a compacted de Bruijn graph
(cDBG)~\cite{Marcus:2014xy,Baier_2015,Beller:2016ab,Chikhi:2015aa,Minkin_2016,Chikhi_2016,almodaresi_et_al:LIPIcs:2017:7657}
from a set of genomes. Basic cDBG does not keep sample information.
\cite{Iqbal:2012aa} proposed colored cDBG with each color represents a sample
or a population. Colored cDBG can be constructed
efficiently~\cite{Muggli_2019,Holley695338}. However, a colored cDBG discards
the chromosomal coordinate and thus disallows the mapping of genomic features.
It often includes connections absent from the input genomes and thus encodes
sequences more than the input. A colored cDBG cannot serve as a
\emph{reference} pangenome graph, either.  deBGA~\cite{Liu:2016ac} addresses
the issue by labeling each unitig with its possibly multiple locations in the
input genome(s). Pufferfish~\cite{Almodaresi:2018aa} further reduces its space
requirement. Nonetheless, given hundreds of human genomes, there will be many
more vertices in the graph and most vertices are associated with hundreds of
labels. Whether deBGA and pufferfish can scale to such datasets remains an open
question. GBWT~\cite{Sir_n_2019} provides another practical solution to storage
and indexing, but no existing tools can practically construct a cDBG for many
human genomes in the GBWT representation.

In addition to cDBG, we can derive a reference pangenome
graph from a single linear multi-sequence alignment (MSA)~\cite{Dilthey_2015,Dilthey_2019}.
It has been used for HLA typing but is not applicable to whole chromosomes when
they cannot be included in a single linear MSA. The third and possibly the most
popular approach to reference graph generation is to call variants from other
sources and then incorporate these variants, often in the VCF format~\cite{Danecek:2011qy}, into
the reference genome as alternative
paths~\cite{Eggertsson:2017aa,Rakocevic_2019,Garrison:2018aa,Sibbesen:2018aa,Biederstedt:2018aa,Eggertsson_2019}.
However, because VCF does not define coordinates on insertions, this approach
cannot properly encode variations on long insertions and is therefore limited
to simple variations. There are no satisfactory solutions to the construction
of reference pangenome graphs.

In this article, we introduce the reference Graphical Fragment Assembly (rGFA)
format to model reference pangenome graphs. We propose and demonstrate an
incremental procedure to construct graphs under this model. The resulting
graphs encode structural variations (SVs) of length 100bp or longer without haplotype
information.  Our implementation, minigraph
(\href{https://github.com/lh3/minigraph}{https://github.com/lh3/minigraph}),
can construct a pangenome graph from twenty human assemblies in three hours.

\section*{Results}

We will first describe a data model for reference pangenome graphs, which
establishes the foundation of this article. We will then present a new
sequence-to-graph mapper, minigraph, and show how this mapper incrementally
constructs a pangenome graph. We will demonstrate the utility of pangenome
graphs with a human graph generated from twenty human haplotypes and a primate
graph generated from four species.

\subsection*{Modeling reference pangenome graphs}

\subsubsection*{Sequence graphs}

There are several equivalent ways to define a sequence graph. In this article,
a \emph{sequence graph} $G(V,E)$ is a bidirected graph. Each vertex $v\in V$ is
associated with a DNA sequence; each edge $e\in E$ has two directions, one for
each endpoint, which leads to four types of edges: forward-forward,
reverse-forward, forward-reverse and reverse-reverse. The directions on an edge
dictate how a sequence is spelled from a walk/path in the graph. Common
assembly graphs, such as the overlap graph, string graph and de Bruijn graph
can all be formulated as sequence graphs.

\begin{figure}[t]
\includegraphics[width=.47\textwidth]{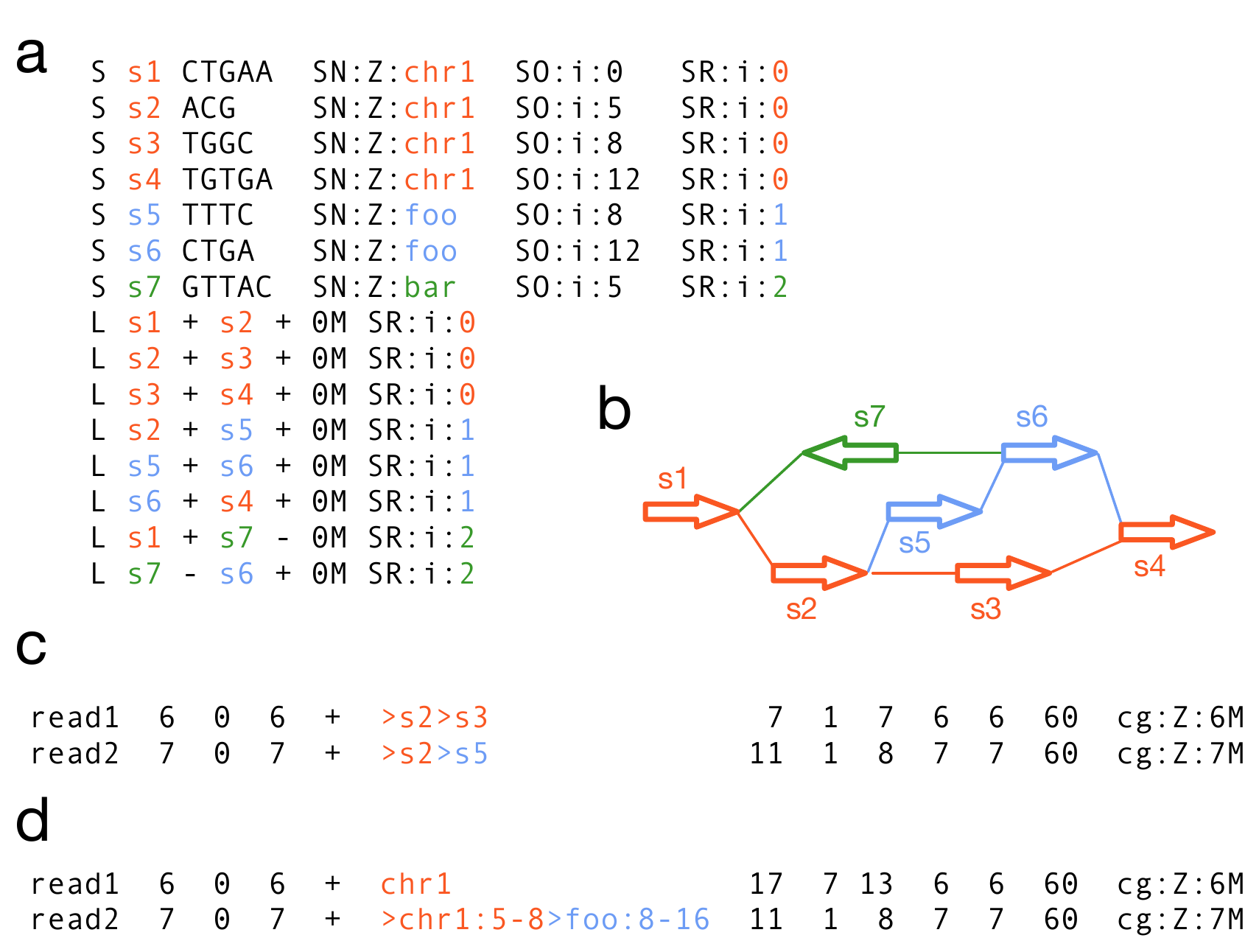}
\caption{\csentence{Example rGFA and GAF formats.} {\bf (a)} Example rGFA
  format. rGFA-specific tags include SN, name of the stable sequence from which
  the vertex is derived; SO, offset on the stable sequence; SR, rank: 0 if the
  vertex or edge is on the linear reference; $>$0 for non-reference.  {\bf (b)}
  Corresponding sequence graph. Each thick arrow represents an oriented DNA
  sequence. {\bf (c)} Example GAF format, using the segment coordinate, for
  reads ``${\tt GTGGCT}$'' and ``${\tt CGTTTCC}$'' mapped to the graph. {\bf
  (d)} Equivalent GAF format using the stable coordinate.}\label{fig:rgfa}
\end{figure}

The Graphical Fragment Assembly (GFA) format~\cite{Li:2016aa} describes
sequence graphs. The core of GFA is defined by the following grammar:

{\footnotesize
\begin{verbatim}

  <GFA>     <- (<segment> | <link>)+
  <segment> <- `S' <segId> <segSeq>
  <link>    <- `L' <segId> [+-] <segId> [+-] <cigar>

\end{verbatim}}

{\flushleft
A line starting with letter ``${\tt S}$'' corresponds to a vertex and a line
starting with ``${\tt L}$'' corresponds
to a bidirected edge. In a de Bruijn graph, we often attach sequences to edges
instead of vertices~\cite{Pevzner:2001vn,Gnerre:2011ys}. To avoid the confusion, in this
article, we also call a vertex as a \emph{segment} and call an edge as a
\emph{link}, following the GFA terminology.  Fig.~\ref{fig:rgfa}a shows an
example GFA that encodes Fig.~\ref{fig:rgfa}b.
}

A sequence graph in the GFA format natively defines a \emph{segment coordinate}
system where each base in the graph is uniquely indexed by a
2-tuple $({\rm segId},{\rm segOffset})$. For example, in
Fig~\ref{fig:rgfa}a, the base at position $({\rm s2},2)$ is ``{\tt G}''.
A major problem with this coordinate is that it is decoupled from linear
annotations and is sensitive to graph transformations. For example, if we split
a segment into two connected segments, the set of sequences spelled from the graph
remains the same, but the segment coordinates will be changed. Due to the
instability of segment coordinate, a basic sequence graph is inadequate for a
reference graph.

\subsubsection*{Reference pangenome graphs}

We propose the reference GFA (rGFA) format to encode reference pangenome graphs.
rGFA is an extension to GFA with three additional tags that indicate the origin
of a segment from linear genomes (Fig.~\ref{fig:rgfa}a). This simple addition
gives us a unique stable coordinate system as an extension to the linear
reference coordinate (e.g. GRCh38). We can pinpoint a position such as
``{\sf chr1:9}'' in the graph and map existing annotations onto the graph. We can
also report a path or walk in the stable coordinate. For example, path
``{\sf s1$\to$s2$\to$s3}'' unambiguously corresponds to ``{\sf
chr1:0-5$\to$chr1:5-8$\to$chr1:8-12}'' or simply ``{\sf chr1:0-12}'' if we
merge adjacent coordinate; similarly, ``{\sf s1$\to$s2$\to$s5$\to$s6}''
corresponds to ``{\sf chr1:0-8$\to$foo:8-16}''. We will formally describe the
path format when introducing the GAF format in the next section.

In rGFA, each segment is associated with one origin. This apparently trivial
requirement in fact imposes a strong restriction on the types of graphs rGFA
can encode: it forbids the collapse of different regions from one sequence,
which would often happen in a cDBG. We consider this restriction an
advantage of rGFA because it requires the graph to have a ``linear'' flavor
intuitively and simplifies the data structure to store the graph.

For simplicity, rGFA disallows overlaps between edges and forbids multiple
edges (more than one edges between the same pair of vertices). These two
restrictions help to avoid ambiguity and reduce the complexity in
implementation. They are not strictly necessary in theory.

\subsubsection*{The Graphical mApping Format (GAF)}

\begin{table}[tb]
\caption{The Graphical mApping Format (GAF)}\label{tab:gaf}
\begin{tabular}{rcp{6cm}}
\hline
Col & Type  & Description \\ \hline
1  & string & Query sequence name \\
2  & int    & Query sequence length \\
3  & int    & Query start coordinate (0-based; closed) \\
4  & int    & Query end coordinate (0-based; open) \\
5  & char   & Strand relative to col. 6 \\
6  & string & Graph path matching regular expression \texttt{/([><][\char94\char92s><]+(:\char92d+-\char92d+)?)+\char124([\char94\char92s><]+)/}\\
7  & int    & Path sequence length \\
8  & int    & Path start coordinate \\
9  & int    & Path end coordinate \\
10 & int    & Number of matching bases in the mapping \\
11 & int    & Number of bases, including gaps, in the mapping \\
12 & int    & Mapping quality (0--255 with 255 for missing) \\ \hline
\end{tabular}
\end{table}

As there are no text formats for sequence-to-graph alignment, we propose a new
Graphical mApping Format (GAF) by extending the Pairwise mApping Format
(PAF)~\cite{Li:2016aa}. GAF is TAB-delimited with each column defined in
Table~\ref{tab:gaf}. Column 6 encodes a path on the graph. It follows the
formal grammar below:

{\footnotesize
\begin{verbatim}

  <path>       <- <stableId> | <orientIntv>+
  <orientIntv> <- (`>' | `<') (<segId> | <stableIntv>)
  <stableIntv> <- <stableId> `:' <start> `-' <end>

\end{verbatim}}

{\flushleft
In this grammar, {\tt <segId>} is a segment identifier on an S-line in rGFA;
{\tt <stableId>} is a stable sequence name at the {\tt SN} tag on the
corresponding S-line. Column 6 can be either a path in the segment coordinate
(Fig.~\ref{fig:rgfa}c) or an equivalent path in the stable coordinate
(Fig.~\ref{fig:rgfa}d). We can merge adjacent stable coordinates if the two
segments are originated from the same stable sequence and the end offset of the
first segment is equal to the start offset of the second segment. For example,
``{\tt >chr1:0-5>chr1:5-8}'' can be simplified to ``{\tt >chr1:0-8}''.
Furthermore, if a path in column 6 is derived from one reference sequence, we
recommend to replace it with the entire reference path on the forward
orientation (e.g. see ``read1'' in Fig.~\ref{fig:rgfa}d). With this convention,
a GAF line is reduced to PAF for a sequence mapped to a reference sequence.
Similar to PAF, GAF also allows optional tags in the SAM-like format. Base
alignment is kept at the {\tt cg} tag.}

Minigraph produces GAF in both the segment and the stable coordinate.
GraphAligner~\cite{Rautiainen810812} produces GAF in the segment coordinate
only, which can be converted to the stable coordinate.

\begin{figure}[t]
\includegraphics[width=.47\textwidth]{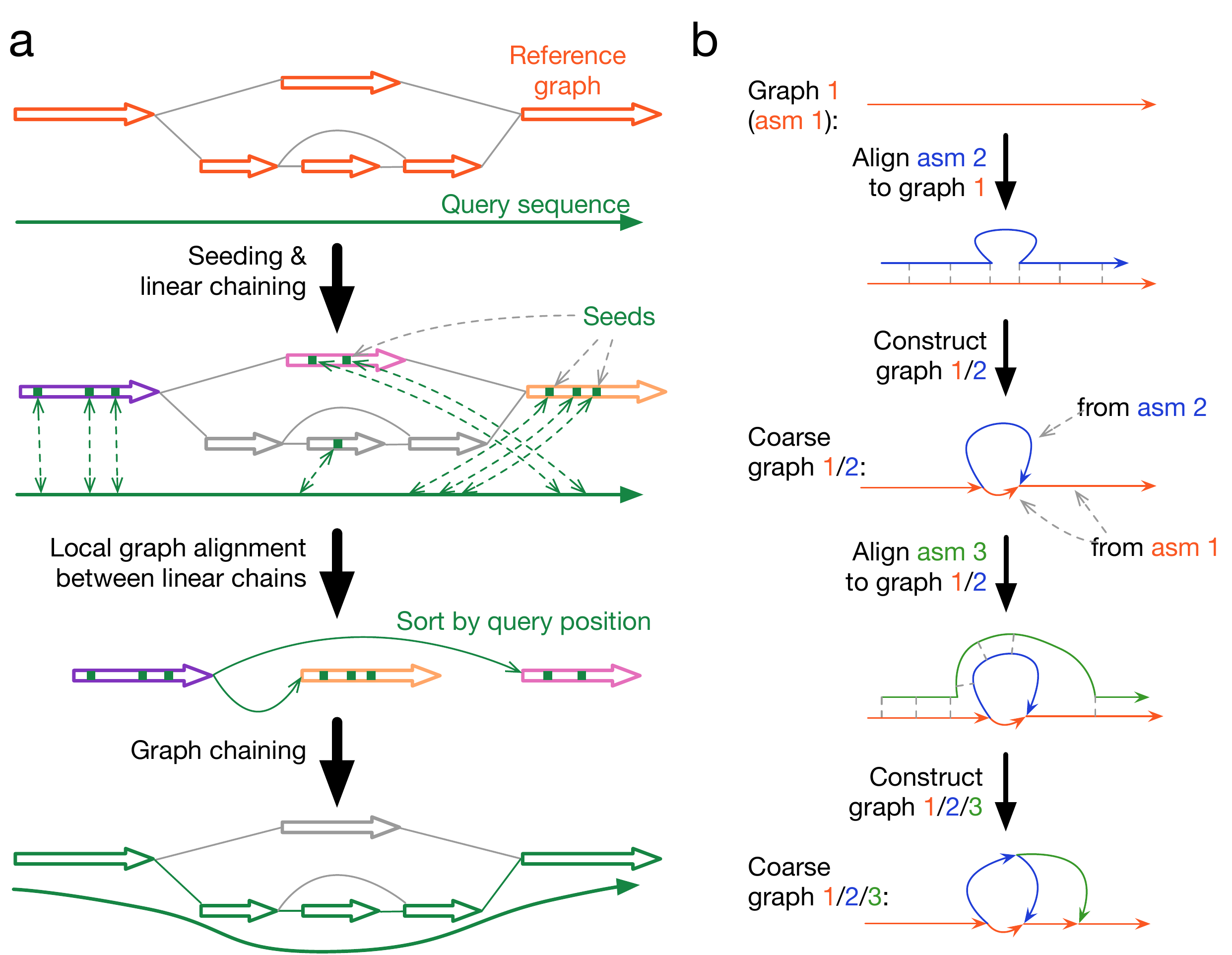}
\caption{\csentence{Minigraph algorithms.} {\bf (a)} Diagram of the minigraph
  mapping algorithm. Minigraph seeds alignments with minimizers, finds good
  enough linear chains, connects them in the graph and seeks the most weighted
  path as a graph chain. {\bf (b)} Diagram of incremental graph construction. A
  graph is iteratively constructed by mapping each assembly to an existing
  graph and augmenting the graph with long poorly mapped sequences in the
  assembly.}\label{fig:mg}
\end{figure}

\subsection*{Sequence-to-graph mapping}

Our incremental graph construction algorithm relies on genome-to-graph
alignment (Fig.~\ref{fig:mg}b). As existing sequence-to-graph
aligners~\cite{Garrison:2018aa,Rautiainen810812} do not work with
chromosome-long query sequences, we adapted minimap2~\cite{Li:2018ab} for our
purpose and implemented minigraph (Fig.~\ref{fig:mg}a). Briefly, minigraph uses
a minimap2-like algorithm to find local hits to segments in the graph, ignoring
the graph topology. It then chains these local hits if they are connected on
the graph, possibly through cycles. This gives the approximate mapping locations. Minigraph does not
perform base-level alignment. This is because the graph we construct encodes
SVs and rarely contains paths similar at the base level. The best mapping is
often clear without base alignment.

\begin{table}[b]
\caption{Performance of sequence-to-graph mapping}\label{tab:mgvga}
\begin{tabular}{lrr}
\hline
& minigraph & GraphAligner \\
\hline
Indexing time (wall-clock sec) & 100 & 589 \\
Mapping time (wall-clock sec) & 79 & 140 \\
Peak RAM (GB)          & 19.5 & 27.2 \\
Percent unmapped reads & 0.5\% & 0\% \\
Percent wrong mappings & 1.7\% & 4.6\% \\
\hline
\end{tabular}
\end{table}

To evaluate the accuracy of minigraph mapping, we simulated PacBio reads from
GRCh38 with PBSIM~\cite{Ono:2013aa} and mapped them to the graph we constructed
in the next section. Table~\ref{tab:mgvga} compares the performance of
minigraph and GraphAligner~\cite{Rautiainen810812} v1.0.10 on 68,857 simulated
reads mapped over 8 CPU threads. Minigraph is faster than GraphAligner and uses
less memory, partly because minigraph does not perform base alignment.

As is shown in Table~\ref{tab:mgvga}, minigraph is more accurate than
GraphAligner. This is counter-intuitive given that GraphAligner does base
alignment. Close inspection reveals that most mismapped reads by minigraph are
mapped to the correct genomic loci but wrong graph paths. On the contrary, most
mismapped reads by GraphAligner are mapped to wrong genomic loci. This suggests
minigraph is better at finding approximate mapping locations but GraphAligner
is better at disambiguating similar graph paths.  Combining the strength of
both could lead to a better graph mapper. We do plan to implement base-level
alignment in minigraph in future.

We have also tried vg v1.21.0. It indexed the same graph in 14.7 wall-clock
hours and mapped the simulated reads in 1.8 hours over 8 threads, tens of times
slower than minigraph and GraphAligner. However, no reads are mapped in the
output. We have not been able to make vg work with our data.

\subsection*{Generating pangenome graphs}

Fig.~\ref{fig:mg}b shows how minigraph constructs a pangenome graph (see
Methods for details). This procedure is similar to multiple sequence alignment
via partial order graph~\cite{Lee_2002} except that minigraph works with cyclic
graphs and ignores small variants. Minigraph only considers SVs of
100bp--100kb in length and ignores SVs in alignments shorter than 100kb.
For each input assembly, it filters out regions covered by two or more primary
alignments longer than 20kb in the assembly. This filter avoids paralogous
regions in a sample and guarantees that graphs generated by minigraph can be
modeled by rGFA.

As a sanity check, we compared minigraph to dipcall
(\href{https://github.com/lh3/dipcall}{https://github.com/lh3/dipcall}) on
calling SVs 100bp or longer from a synthetic diploid sample composed of CHM1
and CHM13~\cite{Li:2018aa}. Given two SV callsets $A$ and $B$, we say a call in
$A$ is \emph{missed} in callset $B$ if there are no calls in $B$ within 1000bp
from the call in $A$. With this criterion, 2.7\% of 14,792 SVs called by
dipcall are missed by minigraph; 6.0\% of 14,932 minigraph SVs are missed by
dipcall. We manually inspected tens of differences in
IGV~\cite{Robinson:2011aa} and identified two causes. First, an INDEL longer
than 100bp called by one caller may be split into two shorter INDELs by the
other caller. There are often more than one smaller SVs around a missed SV
call. Second, dipcall skips regions involving high density of SNPs or involving
both long insertions and long deletions, but minigraph connects these events
and calls SVs in such regions. It tends to call more SVs. Overall, we believe
minigraph and dipcall found similar sets of SVs.

\begin{table}[tb]
\caption{Assemblies used for graph construction}\label{tab:asm}
\begin{tabular}{llll}
\hline
Name & Species & Population  & Accession/Source \\ \hline
CHM1 & Human   & N/A         & GCA\_001297185.1 \\
CHM13 & Human  & N/A         & GCA\_000983455.1 \\
NA12878 & Human & European   & \cite{Garg810341}, phased \\
NA24385 & Human & Jewish     & \cite{Garg810341}, phased \\
PGP1 & Human & N/A           & \cite{Garg810341}, phased \\
NA19240 & Human & African    & GCA\_001524155.4 \\
HG00514 & Human & East Asian & GCA\_002180035.3 \\
HG01352 & Human & American   & GCA\_002209525.2 \\
NA19434 & Human & African    & GCA\_002872155.1 \\
HG02818 & Human & African    & GCA\_003574075.1 \\
HG03486 & Human & African    & GCA\_003086635.1 \\
HG03807 & Human & South Asian& GCA\_003601015.1 \\
HG00733 & Human & American   & GCA\_002208065.1 \\
HG02059 & Human & East Asian & GCA\_003070785.1 \\
HG00268 & Human & European   & GCA\_008065235.1 \\
HG04217 & Human & South Asian& GCA\_007821485.1 \\
AK1     & Human & East Asian & GCA\_001750385.1 \\
Clint   & Chimpanzee &       & GCA\_002880755.3 \\
Susie   & Gorilla &          & GCA\_900006655.3 \\
Kamilah & Gorilla &          & GCA\_008122165.1 \\
Susie   & Orangutan &        & GCA\_002880775.3 \\
\hline
\end{tabular}
\end{table}

\begin{figure*}[htbp]
\includegraphics[width=.95\textwidth]{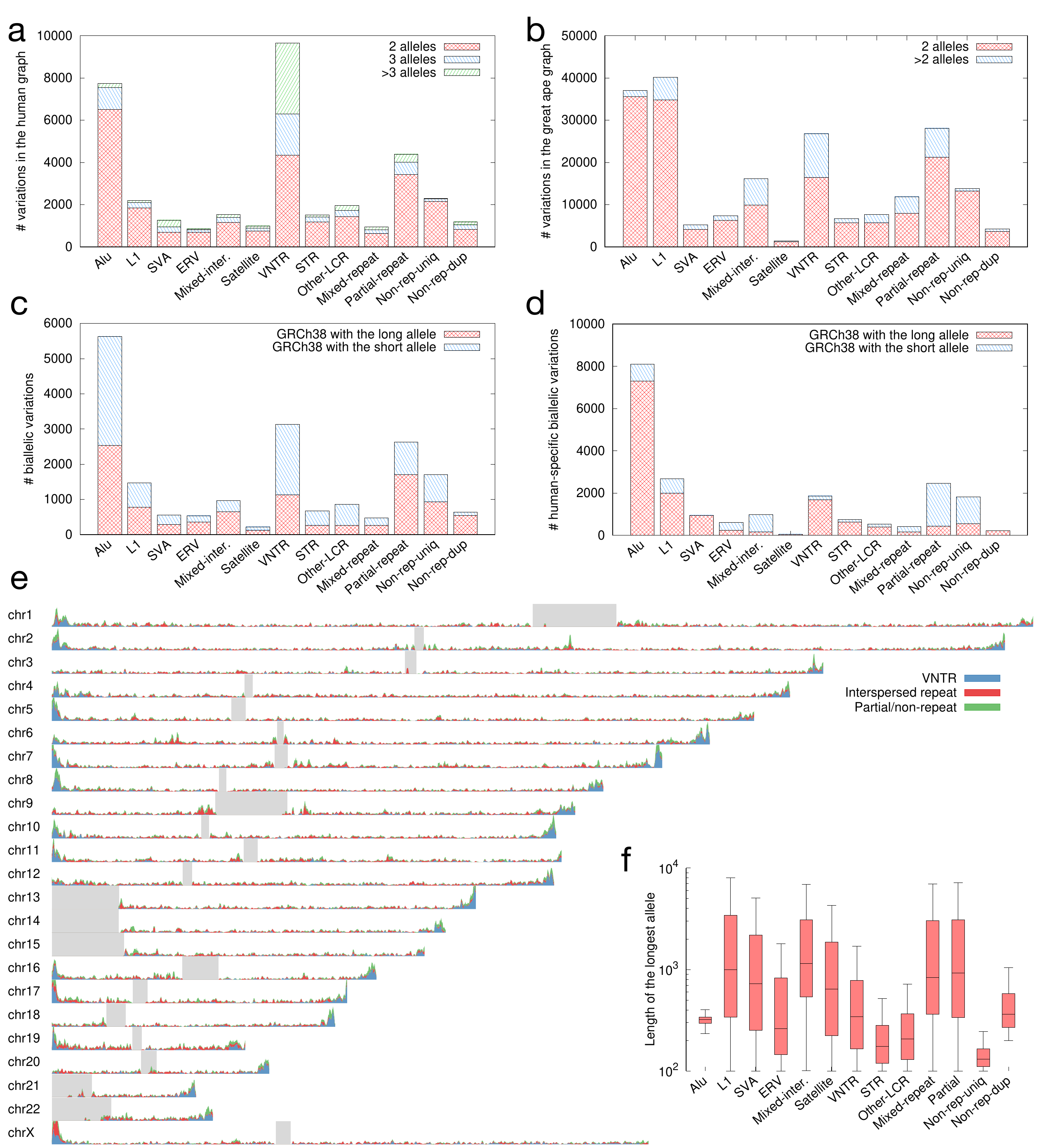}
\caption{\csentence{Characteristics of the human and the great ape graphs.} {\bf
  (a)} Human variations stratified by repeat class and by the number of
  alleles of each variation. The repeat annotation was obtained from the
  longest allele of each variation. VNTR: variable-number tandem repeat, a
  tandem repeat with the unit motif length $\ge$7bp. STR: short random repeat,
  a tandem repeat with the unit motif length $\le$6bp. LCR: low-complexity
  regions. Mixed-inter.: a variation involving $\ge$2 types of interspersed
  repeats. {\bf (b)} Great ape variations stratified by repeat class and by the
  number of alleles. {\bf (c)} Human biallelic variations stratified by repeat
  class and by insertion to/deletion from GRCh38. Both alleles are required to
  be covered in all assemblies. {\bf (d)} Human-specific biallelic variations
  stratified by repeat class and by insertion to/deletion from GRCh38. Red bars
  correspond to insertions to the human lineage. {\bf (e)} Distribution of
  different types of human variations along chromosomes.  {\bf (f)} Boxplot of
  the longest allele length in each repeat class. Outliers are omitted for the
  clarity of the figure.}\label{fig:anno}
\end{figure*}

\subsection*{A human pangenome graph}

Starting with GRCh38, we constructed a human pangenome graph from 20 human
haplotypes or haplotype-collapsed assemblies (Table~\ref{tab:asm}). It took
minigraph 2.7 wall-clock hours over 24 CPU threads to generate this graph. The
peak memory is 98.1GB. The resulting graph consists of 148,618 segments and
214,995 links. It contains 37,332 variations, where a \emph{variation}
denotes a minimal subgraph that has a single source and a single sink with both
segments coming from GRCh38. A path through the bubble between the source and
and the sink represents an \emph{allele}.

Variations in the human graph are enriched with Alus and VNTRs
(Fig.~\ref{fig:anno}a). While interspersed repeats are about evenly distributed
along chromosomes except in the pseudoautosomal regions (Fig.~\ref{fig:anno}e),
VNTRs are enriched towards telomeres~\cite{Audano:2019aa}. It is worth noting
the density of minisatellites is also higher in subtelomeres. If we normalize
the density of VNTRs in the pangenome graph by the density of minisatellites in
GRCh38, the enrichment of VNTRs towards telomeres is still visible but becomes
less prominent. At the same time, repeat-less variations are also enriched
towards the ends of chromosomes (green areas in Fig.~\ref{fig:anno}e),
suggesting subtelomeres tend to harbor SVs anyway. We also
identified 85 processed pseudogenes among these variations.

\begin{figure}
\includegraphics[width=.46\textwidth]{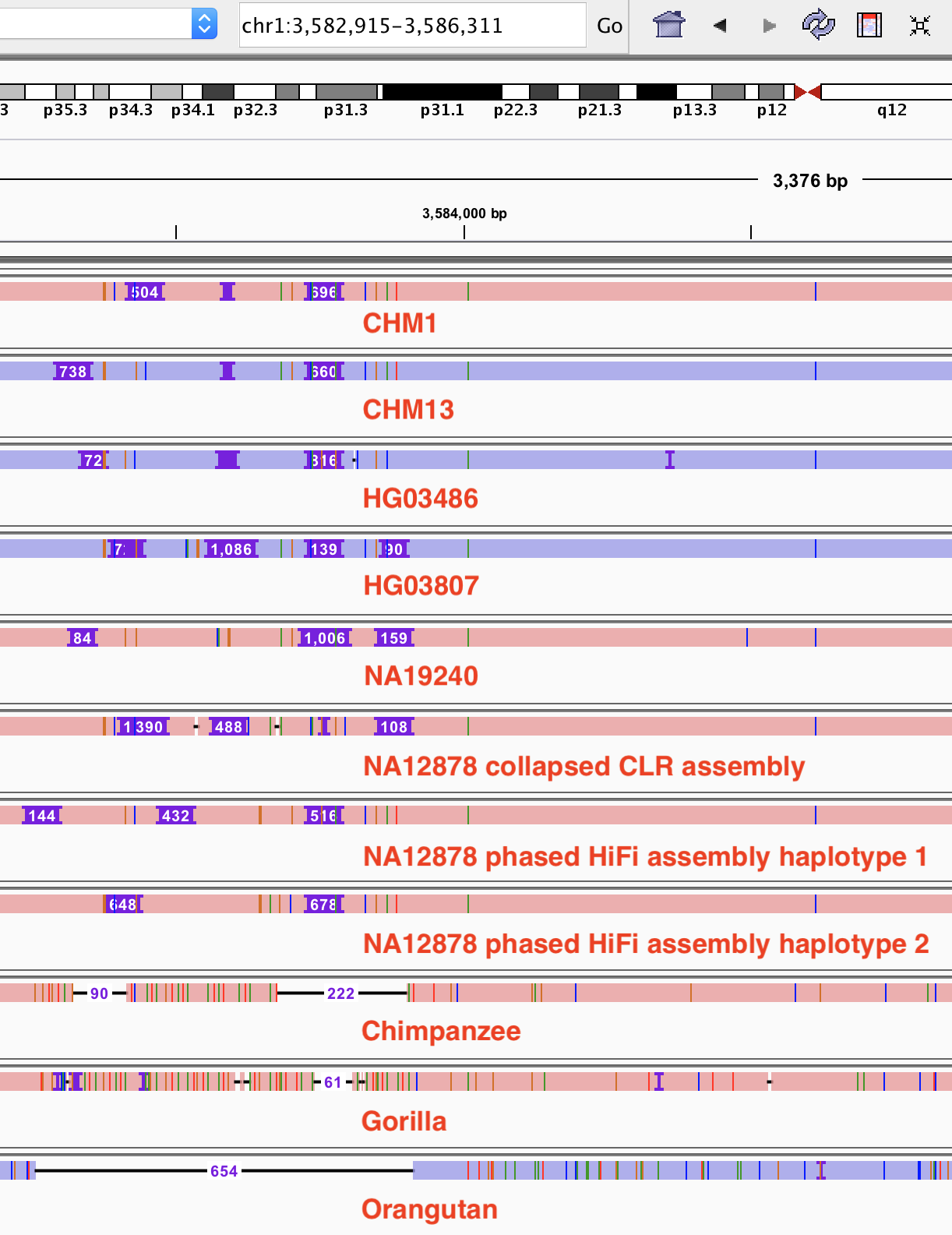}
\caption{\csentence{IGV screenshot of a region enriched with long insertions.}
  Numbers on wide purple bars indicate insertion lengths. CLR: PacBio noisy
  continuous long reads. HiFi: PacBio high-fidelity reads.}\label{fig:igv}
\end{figure}

Another noticeable feature of VNTRs is that over half of VNTR variations are
multiallelic (Fig.~\ref{fig:anno}a). Fig.~\ref{fig:igv} shows a multi-allelic
region composed of VNTRs. We can see many insertions of different lengths. The
two different NA12878 assemblies also disagree with each other, which we often
see around other VNTR loci in NA12878 as well. We have not inspected raw reads
in this particular example, but we tend to believe the disagreement is caused
by local misassemblies rather than somatic mutations. In addition, due to the
multiallelic nature of such VNTRs, the two haplotypes in a human individual are
often different. Assemblies mixing the two haplotypes (aka collapsed
assemblies) may have more troubles in these regions. Multiallelic VNTRs are
hard to assemble correctly.

Multiallelic VNTRs are also hard to align and to call. In Fig.~\ref{fig:igv},
the insertion positions are often different, which could be caused by a few
mutations or sequencing errors. A naive alignment-based SV caller would call a
dozen of low-frequency insertions in this region, which does not reflect these
correlated events. Without base-level alignment, minigraph may
have more troubles with obtaining the optimal alignment in these complex VNTR
regions. Improved data quality, assembly algorithms and graph mapping
algorithms are required to investigate VNTR regions in detail.

\subsection*{A great ape pangenome graph}

We also constructed a great ape pangenome graph from GRCh38, one chimpanzee,
two gorillas and one orangutan (Table~\ref{tab:asm}). This graph contains
206,452 variations, over four times more than the human graph. About half of
variations are originated from orangutan, the species most distant from human.

In the great ape graph, the L1-to-Alu ratio is close to 1:1, much higher than
the ratio in the human graph (Fig.~\ref{fig:anno}b vs Fig.~\ref{fig:anno}a).
This is perhaps correlated with the elevated L1 activity in great
apes~\cite{Mathews:2003aa}. Of retrotransposon-related variations specific to
the human lineage, the overwhelming majority are insertions
(Fig.~\ref{fig:anno}d), which is expected as transpositions lead to insertions
only.  Most human-specific Alu deletions are incomplete and involve ancient Alu
subfamilies. They are likely genomic deletions that happen to hit Alus. In
contrast, the majority of ``partial-repeats'' are deletions from the human
lineage. Two thirds of autosomal insertions in this category are segmental
duplications in GRCh38. In all, minigraph is an efficient tool to study closely
related species.

\subsection*{Blacklist regions from human pangenome graphs}

The human pangenome graph effectively encodes SVs $\ge$100bp
in 20 genomes. These large-scale variations could be a frequent source of
technical artifacts in variant calling with short reads. To test this
hypothesis, we compared short-read SNP calls with vs without regions around SVs
in the pangenome graph.

We constructed a human pangenome graph excluding CHM1 and CHM13, the two
samples used in the SynDip benchmark~\cite{Li:2018aa}, and generated regions
around variations (see Methods), which we call as \emph{blacklist regions},
following the rationale in~\cite{Amemiya:2019aa}.  Blacklist regions is totaled
29.2Mb in length, intersecting 0.7\% of confident regions in
SynDip~\cite{Li:2018aa}; 0.7\% of truth SNPs are contained in blacklist regions
-- true SNPs are not enriched in blacklist regions.

We mapped short reads used in~\cite{Li:2018aa} with minimap2 and called
variants with GATK v4.1.2~\cite{Depristo:2011vn}. This callset
contains 32,879 false positive SNPs, 21\% of which fall in blacklist regions --
false SNP calls are highly enriched in this $<$1\% region of human genome. This
confirms a noticeable fraction of false SNP calls using short reads are
resulted from misalignment involving SVs.

\section*{Discussion}

Based on the GFA assembly format~\cite{Li:2016aa}, we proposed the rGFA format,
which defines a data model for reference pangenome graphs at the same time.
rGFA takes a linear reference genome as the backbone and maintains the
conceptual ``linearity'' of input genomes.

rGFA is not the only pangenome graph model. Vg~\cite{Garrison:2018aa}
encodes a stable sequence with a path through the sequence graph~\cite{10.12688/f1000research.19630.1}. A segment
in the graph may occur on multiple paths, or occur multiple times on one path
if there are cycles in the graph. This way, vg allows different regions in one
chromosome collapsed to one segment. We call such a graph as a collapsed graph. rGFA
cannot encode a collapsed graph. The vg model is thus more general.

In our view, however, the reference pangenome graph should not be a collapsed
graph. In a collapsed graph, the definition of orthology is not clear because
multiple sequences from the same sample may go through the same segment.
Without the concept of orthology, we cannot define variations, either.  In
addition, due to the one-to-many relationship between segments and the
reference genome, it is intricate to derive the stable coordinate of a path in
a collapsed graph. For example, suppose segment {\sf s1} corresponds to two
regions {\sf chr1:100-200} and {\sf chr1:500-600}. To convert a path {\sf
s2$\to$s1$\to$s3} to the stable coordinate, we have to inspect adjacent
segments to tell which {\sf s1} corresponds to; this becomes more challenging
when {\sf s2} and {\sf s3} represent multiple regions in the reference genome.
In contrast, rGFA inherently forbids a collapsed graph and avoids the potential
issues above. This makes rGFA simpler than vg's path model and easier to work
with.

To demonstrate practical applications of rGFA, we developed minigraph to
incrementally generate pangenome graphs. It can generate a graph from 20
genomes in three hours and can scale to hundreds of genomes in future. A
limitation of minigraph is that it does not perform base alignment and may be
confused by similar paths in the graph. We will implement base alignment to
improve the mapping accuracy in such cases. Another limitation of minigraph is
that it is unable to align sequences against a graph encoding all small variants.
Such a graph will be composed of millions of short segments. Not
indexing minimizers across segments, minigraph will fail to seed the initial
linear chains. This limitation can only be resolved by completely changing the
minigraph mapping algorithm. Nonetheless, small variants are easier to
analyze with the standard methods. Incorporating these variants unnecessarily
enlarges the graph, complicates implementations, increases the rate of false
mappings~\cite{Pritt_2018} and reduces the performance of common tasks. There
is also no known algorithm that can construct such a complex graph for hundreds
of human genomes.

Minigraph does not keep track of the sample information as of now. To address
this issue, we are considering to implement colored rGFA, similar to colored de
Bruijn graphs~\cite{Iqbal:2012aa}. In a colored rGFA, a color represents one
sample.  Each segment or link is associated with one or multiple colors,
indicating the sources of the segment or the link. Colors can be stored in an
rGFA tag or in a separate segment/link-by-sample binary
matrix~\cite{Holley695338}. The matrix representation may be more compact given
a large number of samples.

We have shown minigraph can be a fast and powerful research tool to summarize
SVs at the population scale and to study the evolution of closely related
species. A more practical question is how a reference pangenome graph may
influence routine data analysis. Here is our limited view.

We think a critical role a reference graph plays is that it extends the
coordinate system of a linear reference genome. This allows us to annotate
variations in highly diverse regions such as the human HLA and KIR regions. The
existing pipelines largely ignore these variations because most of them cannot
be encoded in the primary assembly of GRCh38.

The extended graph coordinate system further helps to consistently represent
complex SVs. Given multiple samples, the current practice is to call SVs from
individual samples and then merge them. Two subtly different SVs, especially
long insertions, may be called at two distinct locations and treated as
separate events. With the minigraph procedure, the two SVs are likely to
be aligned together as long as they are similar to each other and are
sufficiently different from the reference allele. To some extent, minigraph is
performing multiple sequence alignment with partial order
alignment~\cite{Lee_2002}. This procedure is more robust to different
representations of the same SV than naive merging. When we refer to a SNP, we often use its
chromosomal coordinate such as ``chr1:12345''. We rarely do so for SVs because
their positions are sensitive to alignment and SV callers. The more consistent
SV representation implied by a pangenome graph will help to alleviate the issue
and subsequently facilitate the genotyping of
SVs~\cite{Hickey_2020,Eggertsson_2019,Chen_2019}.

While we believe a reference pangenome graph will make complex variations more
accessible by geneticists and biologists, we suspect a great majority of
biomedical researchers will still rely on a linear reference genome due to the
conceptual simplicity of linear genomes and the mature tool chains developed in
decades. Many analyses such as SNP calling in well behaved regions do not
benefit much from a pangenome representation, either. Nonetheless, a pangenome
reference still helps applications based on linear references. With a graph
reference, we may blacklist regions enriched with SVs that lead to small variant
calling errors.  We may potentially generate ``decoy'' sequences that are
missing from the primary assembly to attract falsely mapped reads away. We may
perform read alignment against a graph, project the alignment to the linear
coordinate and finish the rest of analyses in the linear space. We anticipate a
pangenome reference to supplement the linear reference, not to replace it.

\section*{Conclusions}

Complex human sequence variations are like genomic dark matter: they are
pervasive in our genomes but are often opaque to the assay with the existing
tools. We envision a pangenome graph reference will become an effective
means to the study of these complex variations. We proposed a data model (rGFA),
designed formats (rGFA and GAF) and developed companion tools (minigraph and
gfatools) to demonstrate the feasibility of our vision. Our work is still
preliminary but it is likely to set a starting point to the development of the
next-generation graph-based tools, which may ultimately help us to understand
our genomes better.

\section*{Methods}

\subsection*{The minigraph mapping algorithm}

\subsubsection*{Seeding and linear chaining}
Similar to minimap2, minigraph uses minimizers on segments as seeds. It also
applies a similar chaining algorithm but with different scoring and with a new
heuristic to speed up chaining over long distances. For the completeness of
this article, we will describe part of the minimap2 chaining algorithm here.

\paragraph*{Minimap2-like chaining}
Formally, an \emph{anchor} is a 3-tuple $(x,y,w)$, representing a closed
interval $[x-w+1,x]$ on a segment in the reference graph matching an interval
$[y-w+1,y]$ on the query. Given a list of anchors sorted by $x$, let $f(i)$ be
the maximal chaining score up to the $i$-th anchor in the list. $f(i)$ can be
computed by:
\begin{equation}\label{eq:dp}
f(i)=\max\big\{\max_{i>j\ge1}\{f(j)+\alpha(j,i)-\beta(j,i)\},w_i\big\}
\end{equation}
where $\alpha(j,i)=\min\big\{\min\{y_i-y_j,x_i-x_j\},w_i\big\}$ is
the number of matching bases between anchor $i$ and $j$.
$\beta(j,i)$ is the gap penalty. Let $g_{ji}=|(y_i-y_j)-(x_i-x_j)|$
be the gap length and $d_{ji}=\min\{y_i-y_j,x_i-x_j\}$ be the smaller distance
between the two anchors.  Minigraph uses the following gap cost:
$$
\beta(j,i)=\left\{\begin{array}{ll}
\infty & (g_{ji}>G) \\
c_1\cdot g_{ji} + c_2\cdot d_{ji} + \log_2{g_{ji}} & (0<g_{ji}\le G) \\
0 & (g_{ji}=0)\\
\end{array}\right.
$$
where $G=100000$ in the graph construction mode, $c_1=e^{-dw}$ and
$c_2=0.05\cdot e^{-dw}$. By default, $d=0.01$ is the expected per-base sequence
divergence and $w=19$ is the minimizer length. In comparison, minimap2 applies
$G=5000$, $c_1=0.19$ and $c_2=0$. Minigraph allows much larger gaps between
minimizers and more heavily penalizes gaps.

Solving Eq.~\ref{eq:dp} leads to an $O(n^2)$ algorithm where $n$ is the number
of anchors. This algorithm is slow for large $n$. Minimap2 introduces
heuristics to speed up the computation by approximating this equation. It works
well for minimap2 that only allows small gaps and has base-level alignment as a
fix to chaining errors. However, as minigraph intends to chain much longer
gaps, the minimap2 algorithm occasionally misses the optimal alignment in long
segmental duplications and produces false variations. Minigraph introduces a
new heuristic to speed up chaining.

\begin{figure}[tb]
\centering
\includegraphics[width=.36\textwidth]{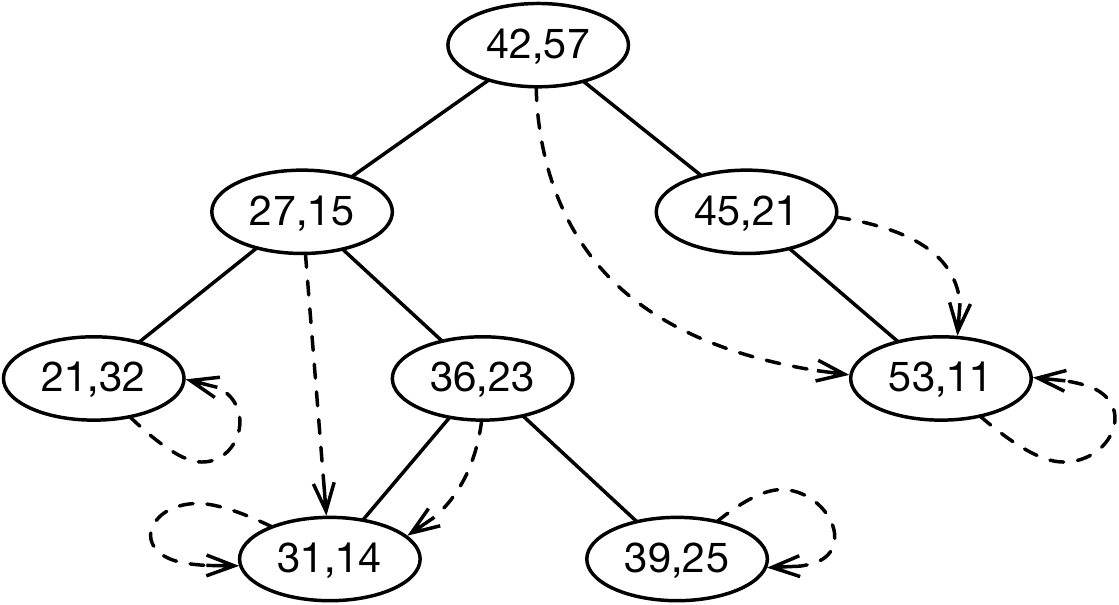}
\caption{\csentence{Implementing 1-dimension Range-Min-Query (RMQ).} Given a
  set of 2-tuples, a binary search tree is built for the first values in the
  tuples. Each node $p$ in the tree is associated with a pointer. The pointer
  points to the node that is in the subtree descended from $p$ and has the
  minimal second value. In this example, ${\rm RMQ}(20,50)=14$.}\label{fig:rmq}
\end{figure}

\paragraph*{Dynamic 1-dimension Range-Min-Query}
Before we move onto the minigraph solution, we will first introduce
Range-Min-Query (RMQ). Given a set of 2-tuples $\{(y_i,s_i)\}$, ${\rm
RMQ}(a,b)$ returns the minimum $s_j$ among $\{s_j:a\le y_j\le b\}$.
We implemented 1-dimension RMQ with a modified AVL tree, a type of balanced
binary search tree (Fig.~\ref{fig:rmq}). When performing ${\rm RMQ}(a,b)$, 
we first find the smallest and the largest nodes within interval $[a,b]$ using
the standard algorithm. In this example, the two nodes are (21,32) and (45,21),
respectively. We then traverse the path between the two nodes to find the
minimum. With a balanced tree structure, we do not need to descend into
subtrees. The time complexity is $O(m\log m)$, where $m$ is the number of nodes
in the tree. We can insert nodes to or delete nodes from the tree while
maintaining the property of the tree. This achieves dynamic RMQ.

\paragraph*{Chaining with a linear gap cost function}
A linear gap cost takes the form of
$\beta'(j,i)=c_1[(y_i-y_j)+(x_i-x_j)]$. Given a list of anchors
$(x_i,y_i,w_i)$ sorted by position $x_i$, let
\begin{equation}\label{eq:dp2}
f'(i)=\max_{\substack{\text{$i>j\ge1$}\\ \text{$x_i-G\le x_j\le x_i-w_i$}\\ \text{$y_i-G\le y_j\le y_i-w_i$}}}\big\{f'(j)+w_j-\beta'(j,i)\big\}
\end{equation}
We can find the optimal $f'(i)$ in $O(n\log n)$ time with
RMQ~\cite{DBLP:conf/wabi/AbouelhodaO03,Otto:2011aa}. To see that, define
$$h'(j)=f'(j)+w_j+c_1(y_j+x_j)$$
The following condition
$$f'(j)+w_j-\beta'(j,i)>f'(k)+w_k-\beta'(k,i)$$
is equivalent to $h'(j)>h'(k)$, independent of $i$. If we maintain ${\rm
RMQ}_i$ as the binary tree that keeps $\{(y_j,-h'(j)):j<i,x_i-G\le x_j\le x_i-w_i\}$, we have
$$
f'(i)=-{\rm RMQ}_i(y_i-G,y_i-w_i)-c_1(x_i+y_i)
$$
This solves Eq.~\ref{eq:dp2} in $O(n\log n)$ time.

\paragraph*{Minigraph linear chaining}
While chaining with a linear gap cost function can be solved efficiently, we
prefer more realistic cost function used in minimap2. In practical
implementation, when we come to anchor $i$, we find the optimal predecessor $j_*$
under the desired gap cost $\beta(j,i)$ for anchors $\{j:j<i,x_i-G'\le
x_j<x_i,y_i-G'\le y_j<y_i\}$, where $G'<G$ is set to 10000 by default.
Meanwhile, we use the RMQ-based algorithm to identify the anchor $j'_{*}$ optimal
under the linear gap cost $\beta'(j,i)$. We choose $j'_*$ as the optimal
predecessor if
$$
f(j_*)+\alpha(j_*,i)-\beta(j_*,i)<f(j'_*)+\alpha(j'_*,i)-\beta(j'_*,i)
$$
This may occasionally happen around long segmental duplications when the
minimap2 heuristic misses the optimal solution. Effectively, minigraph does
thorough search in a small window and approximate search in a large window
using a faster but less sophisticated gap cost function.

\subsubsection*{Graph chaining}

Minigraph generates a set of linear chains $\{L_i\}$ with the procedure above
that completely ignores the graph topology. It then applies another round of
chaining taking the account of the topology.

We say linear chain $L_i$ \emph{precedes} $L_j$, written as $L_i\prec L_j$, if
(1) the ending coordinate of $L_i$ on the query sequence is smaller than the
ending coordinate of $L_j$, and (2) there is a walk from $L_i$ to $L_j$ in the
graph. If there are multiple walks from $L_i$ to $L_j$, minigraph enumerates
the shortest 16 walks and chooses the walk with its length being the closest to
the query distance between $L_i$ and $L_j$.

Given a list of linear chains sorted by their ending coordinates on the query
sequence, let $g(i)$ be the optimal graph chaining score up to linear chain
$L_i$. We can compute $g(i)$ with another dynamic programming:
$$
g(i)=\max\big\{\max_{L_j\prec L_i}\{g(j)+\omega(L_j)-\beta(j,i)\},\omega(L_i)\big\}
$$
where $\beta(j,i)$ is the weight between $L_i$ and $L_j$. As minigraph does not
perform base-level alignment, $\beta(j,i)$ is the same as the gap penalty
function used for linear chaining. $\omega(L_i)$ is the optimal score of $L_i$
computed during linear chaining.

The procedure above has two limitations. First, when computing the weight
between $L_i$ and $L_j$, minigraph largely ignores base sequences and only considers
the distance between them on both the query and the graph. When there are
multiple walks of similar lengths between $L_i$ and $L_j$, minigraph miss the
graph chain that leads to the best base alignment. Although we added a
heuristic by considering 17-mer matches between the query and the graph paths,
we found this heursitc is not reliable in complex regions. Second, minigraph only
enumerates the shortest 16 walks. In complex subgraphs, the optimal walk from
$L_i$ to $L_j$ may not be among them. In the graphs produced by minigraph,
these two limitations only have minor effects, but anyway in future, we
will implement base alignment. We may use the current minigraph algorithm
for easy cases and apply the more expensive base alignment when the current
algorithm potentially fails.

The graph chaining algorithm results in one or multiple graph chains.  A
\emph{graph chain} is a list of anchors $(s_i,x_i,y_i,w_i)$, where
$[x_i-w_i+1,x_i]$ on segment $s_i$ in the graph matches $[y_i-w_i+1,y_i]$ on
the query sequence. A graph chain satisfies the following conditions: if $i<j$,
$y_i<y_j$; if $i<j$ and $s_i=s_j$, we have $x_i<x_j$; if $s_i\not=s_{i+1}$, the
two segments are adjacent on the graph. It is an extension to linear chains.

\subsection*{The minigraph graph generation algorithm}

Using the minimap2 algorithm~\cite{Li:2018ab}, minigraph identifies a set of
\emph{primary chains} that do not greatly overlap with each other on the query
sequence. A region on the query is considered to be \emph{orthogonal} to the
reference if the region is contained in a primary chain longer than 100kb and
it is not intersecting other primary chains longer than 20kb.

Minigraph scans primary chains in orthogonal regions and identifies subregions
where the query subsequences significantly differs from the corresponding
reference subsequences. To achieve that, minigraph computes a score $h_i$ for
each adjacent pair of anchors $(s_i,x_i,y_i,w_i)$ and
$(s_{i+1},x_{i+1},y_{i+1},w_{i+1})$. Let $d^x_i$ be the distance between the
two anchors on the graph and $d^y_i=y_{i+1}-y_i$ be the distance on the query
sequence. $h_i$ is computed as
\begin{equation}\label{eq:hi}
h_i=\left\{\begin{array}{ll}
-10 & \mbox{if $d^x_i=d^y_i\le w_{i+1}$} \\
\eta\cdot\max\{d^x_i,d^y_i\} & \mbox{otherwise}\\
\end{array}\right.
\end{equation}
where $\eta$ is the density of anchors averaged across all primary graph
chains. Define $H(i,j)=\sum_{k=i}^j h_k$. A highly divergent region between the
query and the graph will be associated with a large $H(i,j)$. Minigraph uses
the Ruzzo-Tompa algorithm~\cite{DBLP:conf/ismb/RuzzoT99} to identify all
maximal scoring intervals on list $(h_i)$, which correspond to divergent
regions. In each identified divergent region, minigraph performs base
alignment~\cite{Suzuki:2018aa,Li:2018ab} between the query and the graph
sequences and retains a region if it involves an INDEL $\ge$100bp in length or
a $\ge$100bp region with base-level identity below 80\%. In Eq.~\ref{eq:hi},
-10 is an insensitive parameter due to the downstream filtering. In the end,
minigraph augments the existing graph with identified variations
(Fig.~\ref{fig:mg}b).

\subsection*{Annotating variations}

We applied RepeatMasker~\cite{Tarailo-Graovac:2009aa} v1.332 to classify
interspersed repeats in the longest allele sequence of each variation.
RepeatMasker is unable to annotate VNTRs with long motifs. It also often
interprets VNTRs as impure STRs. Therefore, we did not use the RepeatMasker
VNTR or STR annotations directly. Instead, we combined RepeatMasker and
SDUST~\cite{Morgulis:2006aa} results to collect low-complexity regions (LCRs).
We identified pure tandem repeats composed of a motif occurring twice or more
(implemented in
\href{https://github.com/lh3/etrf}{https://github.com/lh3/etrf}). An LCR is
classified as VNTR if 70\% of the LCR is VNTR; similarly, an LCR is classified
as STR if 70\% is STR; the rest are classified as ``Other-LCR'' in
Fig.~\ref{fig:anno}. The annotation script is available in the minigraph GitHub
repository.

\subsection*{Creating blacklist regions}

For each variation in the graph, we extend its genomic interval on GRCh38 by
50bp from each end. We name this set of intervals as $I_0$. We align sequences
inserted to GRCh38 against GRCh38 with ``minimap2 -cxasm20 -r2k'' and filter
out alignments with mapping quality below 5. Let $I(a,b)$ be the set of GRCh38
intervals that are contained in alignments with identity between $a$ and $b$.
The blacklist regions are computed by $I_0\cup I(0,0.99)\setminus I(0.998,1)$,
where ``$\cup$'' denotes the interval union operation and ``$\setminus$''
denotes interval subtraction.


\begin{backmatter}

\section*{Competing interests}
The authors declare that they have no competing interests.

\section*{Author's contributions}
HL conceived the project, developed minigraph and drafted the manuscript.
XF did the pseudogene analysis. CC helped with RepeatMasker annotation.
All authors helped to revise the manuscript.

\section*{Acknowledgements}
We are grateful to Benedict Paten and Erik Garrison for discussions on
pangenome graphs. We thank minigraph users who have suggested features and
helped to fix various issues.

\section*{Funding}
This work is supported by National Institutes of Health (NIH) grant
U01HG010961 and R01HG010040.

\section*{Availability of data and materials}
Minigraph is openly available at
\href{https://github.com/lh3/minigraph}{https://github.com/lh3/minigraph}.
This repository also includes the script to convert from the segment coordinate
to the stable coordinate, to annotate variations and to generate blacklist
regions from the graph. The companion gfatools is available at
\href{https://github.com/lh3/gfatols}{https://github.com/lh3/gfatools}. The
human and the great ape graphs are hosted at
\href{ftp://ftp.dfci.harvard.edu/pub/hli/minigraph/}{ftp://ftp.dfci.harvard.edu/pub/hli/minigraph/}.
The NA12878, NA24385 and PGP1 phased assemblies were downloaded from
\href{ftp://ftp.dfci.harvard.edu/pub/hli/whdenovo/}{ftp://ftp.dfci.harvard.edu/pub/hli/whdenovo/}.
Other assemblies are all available from GenBank.

\bibliographystyle{bmc-mathphys}
\bibliography{minigraph}









\end{backmatter}
\end{document}